# Nine New Variable Stars in Cygnus and Variability Type Determination of [Wm2007] 1176

**Riccardo Furgoni**

*Keyhole Observatory, Via Fossamana 86, S. Giorgio di Mantova (MN), Italy*

*and*

*AAMN Gorgo Astronomical Observatory MPC 434, S. Benedetto Po (MN), Italy; riccardo.furgoni@alice.it*

**Abstract** I report the discovery of nine new variable stars in Cygnus: five pulsating (VSX J192319.8+280832, VSX J192405.8+280352, VSX J192220.7+275518, VSX J192304.4+280231, VSX J192255.1+274744) and four eclipsing (VSX J192252.4+280217, VSX J192251.4+280456, VSX J192226.0+281019, VSX J192524.9+275342). The variability type of the variable star [WM2007] 1176, that was considered in literature a possible RRC, was found to be a W UMa variable with an obvious O'Connell effect.

**1. Introduction**

A photometric campaign aimed at the study of the variable star [WM2007] 1176 discovered in the CCD/Transit Instrument Survey (CTI-I; Wetterer and McGraw 2007) database was carried out in order to accurately determine its type. This variable was in fact considered in the literature RRC but the assignment was uncertain. Moreover, all the light curves of the stars present in the imaging field between magnitude 9 V and 16 V have been inspected visually: nine of these were found to present a periodic variability. This study highlights how the simple visual inspection of the light curves is a good way to search for new variables, especially for those who have light variations of the order of a few hundredths of a magnitude and low SNR. In the latter case, the creation of the classic diagram "mean magnitude versus RMS-scatter" does not often immediately reveal variable stars.

**2. Instrumentation used**

The data were obtained with a Celestron C8 Starbright, a Schmidt-Cassegrain optical configuration with aperture of 203 mm and central obstruction of 34%. The telescope was positioned at coordinates 45° 12' 33" N, 10° 50' 20" E (WGS84) at the Keyhole Observatory, a new roll-off roof structure managed by the author and shown in Figure 1.

The telescope is equipped with a focal reducer Baader Planetarium Alan

Gee II able to bring the focal length from 2030 mm to 1396 mm, and the focal ratio to $f$ 6.38 from the original $f$ 10.

The pointing is maintained with a Syntha NEQ6 mount with software synscan 3.27, guided using a Baader Vario Finder telescope equipped with a Barlow lens capable of bringing the focal length of the system to 636 mm and focal ratio of $f$ 10.5.

The guide camera is a Magzero MZ-5 with Micron MT9M001 monochrome sensor equipped with an array of 1280X1024 pixel. The size of the pixels is 5.2 µm × 5.2 µm for a resulting sampling of 1.68 arcsec/pixel.

The CCD camera is an SBIG ST8300m with monochrome sensor Kodak KAF8300 equipped with an array of 3352 × 2532 pixels. The pixels are provided with microlenses for improving the quantum efficiency, and the size of the pixels is 5.4 µm × 5.4 µm for a resulting sampling of 0.80 arcsec/pixel. The camera has a resolution of 16 bits with a gain of 0.37e-/ADU and a full-well capacity of 25,500 electrons. The dark current is 0.02e-/pixel/sec at a temperature of –15° C. The typical read noise is 9.3e-RMS.

This camera has some unusual features compared to a CCD normally used in photometry: a high gain and a characteristic peak sensitivity in the V passband. Most of the CCDs normally used in photometry are much more sensitive in the R passband.

The camera is equipped with a 1,000× antiblooming: after exhaustive testing it has been verified that the zone of linear response is between 1,000 and 20,000 ADU, although up to 60,000 ADU the loss of linearity is less than 5%. The CCD is equipped with a single-stage Peltier cell T = 35 ± 0.1° C which allows cooling at a stationary temperature.

**3. Data collection**

The observed field is centered at coordinates R.A. (J2000) $19^h 23^m 52^s$, Dec. +27° 57' 56" and its dimensions are 44.6' × 33.7' with a position angle of 358°.

The observations were performed with the CCD at a temperature of –5° C in binning 1 × 1. The exposure time was 55 seconds with a delay of 1 second between the images and an average download time of 11 seconds per frame. The observations were carried out without the use of photometric filters to maximize the signal-to-noise ratio. The spectral sensitivity of the CCD, as shown in Figure 2, is maximized at a wavelength of 540 nm, making the data more compatible with a magnitude CV (Clear Filter – Zero Point V).

The observations were conducted over seven nights as presented in Table 1.

The CCD control program was Software Bisque's ccdsoft V5. Once the images were obtained, the calibration frames were taken for a total of 100 dark of 55 seconds at –5° C, 200 darkflat of 2 seconds at –5° C, 100 flat of 2 seconds at –5° C. The darkflats and darks were taken only at the first observation session and were used for all other sessions. The flats were taken for each session as the position of the CCD camera as well as the focus point could be varied slightly.

The calibration frames were combined with the method of the median and the masterframe obtained was then used for the correction of the images taken. All images were then aligned and an astrometric reduction made to implement the astrometrical coordinate system WCS in the FITS header. These operations were conducted entirely through the use of software MAXIMDL V5.18 made by Diffraction Limited.

**4. Deriving a magnitude in unfiltered photometry**

Unfortunately it is not always possible to have AAVSO comparison star sequences in all fields of view. Furthermore, by performing unfiltered differential photometry it is crucial to use stars of color as close as possible to the star to be measured in order to avoid a different atmospheric extinction degrading the data obtained, especially when the photometric sessions are long and involve different air masses.

I used this approach to provide a CV magnitude (unfiltered photometry with V-zero point) as accurate and as close to the likely real values for the V magnitude as possible. From an estimate made in this photometric campaign and in others made for the same purpose, it has been observed that in the presence of comparison stars of color similar to the star to be measured, the results obtained with an unfiltered CCD with a peak sensitivity in the V band allow you to get closer to the real values V with deviations of only a few hundredths of magnitude than those obtained using a photometric filter.

Proof of this is evident by analyzing the difference between the average magnitude of the δ Scuti stars discovered in this research campaign with respect to V magnitudes obtained from the magnitude r' in the Carlsberg Meridian Catalog 14 (Copenhagen Univ. Obs. 2006), as described later in this work. Only δ Scuti stars were used for comparison because of the minimal amplitude of their variation. For the other stars the comparison does not make sense since we do not know at what time of their phase CMC14 made a measurement. The results are presented in Table 2.

In detail, the method used was as follows: when by a first rough inspection of the light curve a star was detected as a possible new variable, its color was determined by using the value of J–H on the magnitudes provided by 2MASS All-Sky Catalog of Point Sources (Cutrie *et al.* 2003). The VizieR catalog, operated by CDS in Strasbourg, was searched for stars with a J–H index as close as possible to the candidate variable, a shorter angular distance, and adequate SNR. A comparison star and a check star were identified for each variable in order to provide a V magnitude as correct as possible for the comparison star. Existing photometric data were used when possibile or V magnitudes were derived. The sources used in order of preference were:

*APASS* (Henden *et al.* 2012): this survey is certainly one of the best available for star magnitude determination in different passbands. Unfortunately, in the observed field there are not APASS data available.

*ASAS3* (Pojmański 2002): this survey usually provides precise photometric results for the V passband. The short focal length of the telescopes used, however, produces significant blending problems and sometimes even if a star is listed in a specific position, the data provided regard different stars mixed together. When this phenomenon was evident or suspected (often in the case of a crowded field such as this one), it was decided to not use the ASAS3 data for reference.

*TASS MARK IV* (Droege *et al.* 2006): this survey provides good photometric results for the V passband. The short focal length of the telescopes used produces problems similar to those found in ASAS3. The data were used in this case with the same criterion as those relating to ASAS3.

*UCAC3* (Zacharias *et al.* 2010) and *CMC14* (Copenhagen Univ. Obs. 2006): these surveys provide photometry in different bands with respect to V passband. Transformation equations were used, however, to derive a magnitude in the V passband with reasonable accuracy. The problems of blending are much more rare. The equations used are:

$$V_{Johnson-Cousins} = 0.6278 (J_{2MASS} - K_{2MASS}) + 0.9947 \, r'CMC14 \quad (9 < r' < 16) \quad (1)$$

$$V_{Johnson-Cousins} = 0.531 (J_{2MASS} - K_{2MASS}) + 0.9060 \, fmagUCAC3 + 0.95 \quad (2)$$

Equation (1) for CMC14 is described in Dymock and Myles (2009). Equation (2) for UCAC3 is described in Pavlov (2009).

## 5. Magnitude determination and period calculation

The star's brightness was measured with MAXIMDL V5.18 software, using the aperture ring method. With a FWHM of the observing sessions at times arrived at values of 4" it was decided to choose values providing an adequate signal-to-noise ratio and the certainty of being able to properly contain the whole flux sent from the star. I have used the following apertures: Aperture radius, 12 pixels; Gap width, 2 to 28 pixels; Annulus thickness, 8 pixels.

Before proceeding further in the analysis, the time of the light curves obtained was heliocentrically corrected (HJD) in order to ensure a perfect compatibility of the data with observations carried out even at a considerable distance in time. The determination of the period was calculated using the software PERIOD04 (Lenz and Breger 2005), using a Discrete Fourier Transform (DFT). The average zero-point (average magnitude of the object) was subtracted from the dataset to prevent the appearance of artifacts centered at frequency 0.0 of the periodogram. The calculation of the uncertainties was carried out with PERIOD04 using the method described in Breger *et al.* (1999).

### 6. VSX J192319.8+280832

Position (UCAC4): R.A. (J2000) = $19^h$ $23^m$ $19.895^s$, Dec. (J2000) = +28° 08' 32.27"

Cross Identification: 2MASS J19231989+2808322; GSC 02136-00868; UCAC3 237-177327; UCAC4 591-080941

Variability Type: δ Sct

Magnitude: 13.23 CV (0.03 magnitude amplitude)

Main Period: 0.14296(2) d

Secondary Period: 0.064818(4) d

Epoch Main Period: 2456105.298(2) HJD

Epoch Secondary Period: 2456105.4085(7) HJD

Comparison Star: 2MASS J19233744+2808289

Comp. Star Magnitude: 12.685 V (UCAC3 derived V magnitude)

Check Star: 2MASS J19225994+2809312

Finding chart, phase plots, and Fourier spectrum are shown in Figures 3, 4, 5, and 6.

About this star and all other pulsating stars discovered in this work I want to clarify that they are to the limit of the detection capability of the instrumentation and the only information obtainable from these data is that there exists a periodic pulsation. Considering the color index of these stars, the amplitude of the pulsation, and the duration of the period, there is no possibility that they are anything other than δ Scuti type variables.

### 7. VSX J192405.8+280352

Position (UCAC4): R.A. (J2000) = $19^h$ $24^m$ $05.820^s$, Dec. (J2000) = +28° 03' 52.06"

Cross Identification: 2MASS J19240582+2803520; GSC 02133-00013; TYC 2133-13-1; UCAC3 237-178258; UCAC4 591-081338

Variability Type: δ Sct

Magnitude: 11.47 CV (0.01 magnitude amplitude)

Period: 0.0365621(12) d

Epoch: 2456105.4662(4) HJD

Comparison Star: 2MASS J19243630+2808488

Comp. Star Magnitude: 12.289 V (UCAC3 derived V magnitude)

Check Star: 2MASS J19233463+2810322

Finding chart, phase plot, and Fourier spectrum are shown in Figures 7, 8, and 9.

## 8. VSX J192252.4+280217

Position (UCAC4): R.A. (J2000) = $19^h 22^m 52.436^s$, Dec. (J2000) = +28° 02' 17.32"

Cross Identification: 2MASS 19225244+2802175; CMC14 J192252.4+ 280217; UCAC3 237-176692; UCAC4 591-080622; USNO-B1.0 1180-0416101

Variability Type: eclipsing binary β Per type

Magnitude: maximum 15.37 CV; minimum 17.7: CV

Period: 2.2957(8) d

Epoch: 2456135.468(6) HJD

Comparison Star: 2MASS J19231832+2808483

Comp. Star Magnitude: 12.895 V (UCAC3 derived V magnitude)

Check Star: 2MASS J19222259+2756428

The minimum in CV magnitude was calculated using a binning method due to the very low SNR of the star. Finding chart and phase plot are shown in Figures 10 and 11.

## 9. VSX J192251.4+280456

Position (UCAC4): R.A. (J2000) = $19^h 22^m 51.419^s$, Dec. (J2000) = +28° 04' 56.51"

Cross Identification: 2MASS J19225142+2804564; CMC14 J192251.4+ 280456; UCAC3 237-176666; UCAC4 591-080609; USNO-B1.0 1180-0416062

Variability Type: eclipsing binary β Lyr type

Magnitude: maximum 13.53 CV; minimum 13.83 CV (secondary minimum 13.75 CV)

Period: 0.3562496(92) d

Epoch: 2456126.4155(6) HJD

Comparison Star: 2MASS J19230552+2802117

Comp. Star Magnitude: 13.492 V (Mean value of UCAC3 derived V magnitude and CMC14 derived V magnitude)

Check Star: 2MASS J19225414+2759330

Finding chart, phase plot, and Fourier spectrum are shown in Figures 12, 13, and 14.

### 10. VSX J192226.0+281019

Position (UCAC4): R.A. (J2000) = $19^h 22^m 26.026^s$, Dec. (J2000) = +28° 10' 19.32"

Cross Identification: 2MASS J19222602+2810193; CMC14 J192226.0+ 281019; UCAC3 237-176164; UCAC4 591-080337; USNO-B1.0 1181-0395081

Variability Type: eclipsing binary W UMa type

Magnitude: Max 13.04 CV – Min 13.20 CV (possible presence of O'Connell effect)

Period: 0.2852148(26) d

Epoch: 2456135.3331(2) HJD

Comparison Star: 2MASS J19221496+2803136

Comp. Star Magnitude: 12.126 V (Mean value of ASAS3 Vmag, TASS Mark IV Vmag, UCAC3 derived Vmag and CMC14 derived Vmag)

Check Star: 2MASS J19223002+2805008

Finding chart, phase plot, and Fourier spectrum are shown in Figures 15, 16, and 17.

### 11. VSX J192524.9+275342

Position (UCAC4): R.A. (J2000) = $19^h 25^m 24.927^s$, Dec. (J2000) = +27° 53' 42.95"

Cross Identification: 2MASS J19252492+2753429; CMC14 J192524.9+275342; UCAC3 236-180684; UCAC4 590-083884; USNO-B1.0 1178-0464728

Variability Type: eclipsing binary W UMa type

Magnitude: maximum 14.23 CV; minimum 14.40 CV

Period: 0.518611(13) d

Epoch: 2456126.425(1) HJD

Comparison Star: 2MASS J19250662+2744573

Comp. Star Magnitude: 12.477 V (Mean value of ASAS3 V magnitude, TASS Mark IV V magnitude, UCAC3 derived V magnitude and CMC14 derived V magnitude)

Check Star: 2MASS J19251846+2752372

Finding chart, phase plot, and Fourier spectrum are shown in Figures 18, 19, and 20.

### 12. VSX J192220.7+275518

Position (UCAC4): R.A. (J2000) = $19^h$ $22^m$ $20.781^s$, Dec. (J2000) = +27° 55' 17.97"

Cross Identification: 2MASS J19222078+2755179; CMC14 J192220.7+275517; UCAC3 236-176830; UCAC4 590-082049; USNO-B1.0 1179-0436072

Variability Type: δ Sct

Magnitude: 12.24 CV (0.02 magnitude amplitude)

Period: 0.075664(5) d

Epoch: 2456135.4681(7) HJD

Comparison Star: 2MASS J19232689+2759395

Comp. Star Magnitude: 11.427 V (Mean value of ASAS3 V magnitude, TASS Mark IV V magnitude, UCAC3 derived V magnitude and CMC14 derived V magnitude)

Check Star: 2MASS J19222757+2754528

Finding chart, phase plot, and Fourier spectrum are shown in Figures 21, 22, and 23.

### 13. VSX J192304.4+280231

Position (UCAC4): R.A. (J2000) = $19^h$ $23^m$ $04.402^s$, Dec. (J2000) = +28° 02' 31.85"

Cross Identification: 2MASS J19230439+2802318; CMC14 J192304.4+280231; UCAC4 591-080752; USNO-B1.0 1180-0416528

Variability Type: δ Sct

Magnitude: 13.28 CV (0.02 magnitude amplitude)

Period: 0.113080(12) d

Epoch: 2456126.4211(12) HJD

Comparison Star: 2MASS J19232689+2759395

Comp. Star Magnitude: 11.462 V (ASAS3 V magnitude)

Check Star: 2MASS J19222757+2754528

Finding chart, phase plot, and Fourier spectrum are shown in Figures 24, 25, and 26.

### 14. VSX J192255.1+274744

Position (UCAC4): R.A. (J2000) = $19^h 22^m 55.127^s$, Dec. (J2000) = +27° 47' 44.62"

Cross Identification: 2MASS J19225512+2747445; CMC14 J192255.1+274744; UCAC3 236-177521; UCAC4 589-080251; USNO-B1.0 1177-0462865

Variability Type: δ Sct

Magnitude: 12.02 CV (0.02 magnitude amplitude)

Period: 0.069443(5) d

Epoch: 2456125.3931(8) HJD

Comparison Star: 2MASS J19232689+2759395

Comp. Star Magnitude: 11.462 V (ASAS3 V magnitude)

Check Star: 2MASS J19222757+2754528

Finding chart, phase plot, and Fourier spectrum are shown in Figures 27, 28, and 29.

### 15. [WM2007] 1176

Position (UCAC4): R.A. (J2000) = $19^h 24^m 03.116^s$, Dec. (J2000) = +28° 02' 27.64"

Cross Identification: 2MASS J19240312+2802276; CMC14 J192403.1+280227; UCAC3 237-178198; UCAC4 591-081311

Variability Type: eclipsing binary W UMa type

Magnitude: maximum 13.47 V; minimum 13.83 V (secondary maximum 13.50 V), presence of O'Connell effect

Period: 0.613472(1) d

Epoch: 2456126.4101(7) HJD

Comparison Star: 2MASS J19232689+2759395

Comp. Star Magnitude: 11.462 V (ASAS3 V magnitude)

Check Star: 2MASS J19243203+2804542

The magnitude range in the V passband was determined using CTI-I V magnitude data. The period is based on the best fit for both FRIC dataset and CTI-I dataset. The best period for only CTI-I dataset is 0.613470 day: therefore a possibile variation of the period occurred over time. Finding chart, phase plot, and Fourier spectrum are shown in Figures 30, 31, and 32.

## 16. Acknowledgements

I wish to thank Dr. Patrizia Caviezel for her contribution in discovering the EA-type variable VSX J192252.4+280217. This work has made use of the VizieR catalogue access tool, CDS, Strasbourg, France, and the International Variable Star Index (VSX) operated by the AAVSO. This work has made use of NASA's Astrophysics Data System and data products from the Two Micron All Sky Survey, which is a joint project of the University of Massachusetts and the Infrared Processing and Analysis Center/California Institute of Technology, funded by the National Aeronautics and Space Administration and the National Science Foundation. This work has made use of *ASAS3 Public Catalog*, *Carlsberg Meridian Catalog 14*, the *Third U.S. Naval Observatory CCD Astrograph Catalog* (UCAC3), and the *Fourth U.S. Naval Observatory CCD Astrograph Catalog* (UCAC4).

Table 1. Dates and times of observations.

| Date dd-mm-yyyy | UTC Start hh:mm:ss | UTC End[1] hh-mm-ss | Useful Exposures |
|---|---|---|---|
| 26-06-2012 | 21:26:38 | 00:40:17 | 170 |
| 03-07-2012 | 20:30:24 | 00:45:47 | 221 |
| 16-07-2012 | 20:22:05 | 00:30:31 | 217 |
| 17-07-2012 | 20:23:23 | 00:49:15 | 232 |
| 18-07-2012 | 20:20:24 | 23:00:25 | 140 |
| 26-07-2012 | 20:12:51 | 01:04:41 | 255 |
| 02-08-2012 | 19:59:13 | 22:14:59 | 118 |

[1] When the UTC end exceeds 23:59 it refers to the day after the start of exposures.

Table 2. Average CV magnitude (CV=Clear passband with V zero-point) of the Delta Scuti stars discovered in this research campaign compared to the V magnitude derived from the magnitude r' in the *Carlsberg Meridian Catalog 14*.

| Star | CMC14 r' mag. | 2MASS Jmag. | 2MASS Kmag. | CMC 14 Vmag. | CV mag Furgoni | Abs. Mag. Difference |
|---|---|---|---|---|---|---|
| VSX J192319.8+280832 | 13.13 | 12.08 | 11.74 | 13.28 | 13.23 | 0.05 |
| VSX J192405.8+280352 | 11.50 | 10.76 | 10.62 | 11.53 | 11.47 | 0.06 |
| VSX J192220.7+275518 | 12.15 | 11.30 | 11.09 | 12.21 | 12.24 | 0.03 |
| VSX J192304.4+280231 | 13.13 | 12.15 | 11.87 | 13.23 | 13.28 | 0.05 |
| VSX J192255.1+274744 | 11.87 | 10.99 | 10.77 | 11.95 | 12.02 | 0.07 |

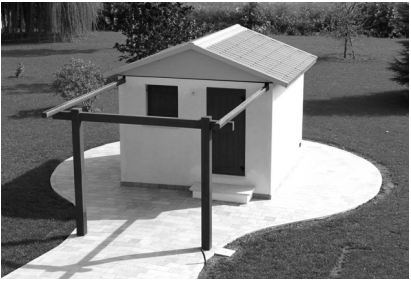

Figure 1. The Keyhole Observatory.

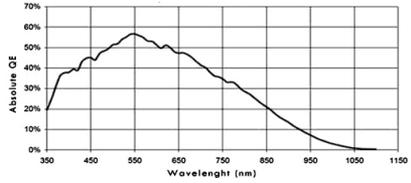

Figure 2. Quantum efficiency of the sensor KAF8300 Monochrome with microlens.

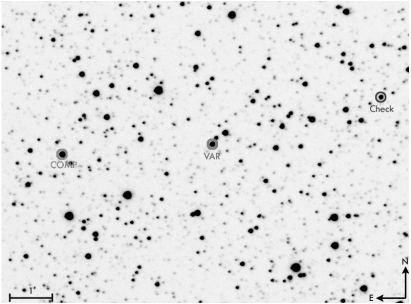

Figure 3. Finding chart of VSX J192319.8+280832. Scale marker indicates one arc minute.

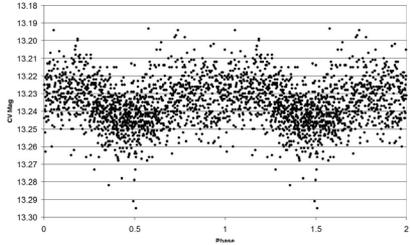

Figure 4. Main period phase plot of VSX J192319.8+280832.

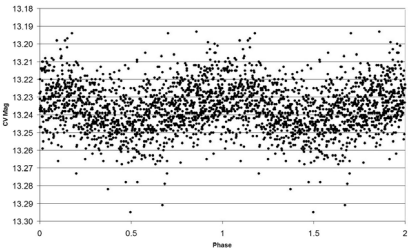

Figure 5. Secondary period phase plot of VSX J192319.8+280832.

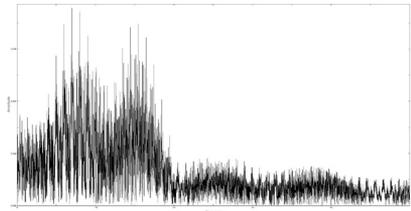

Figure 6. Fourier spectrum of VSX J192319.8+280832.

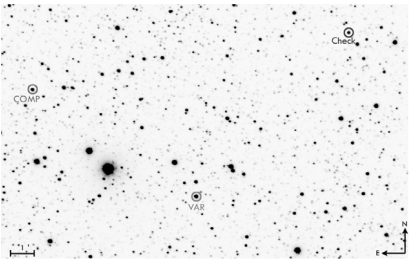

Figure 7. Finding chart of VSX J192405.8+280352.

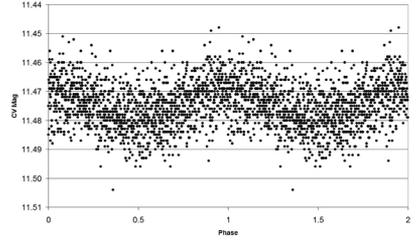

Figure 8. Phase plot of VSX J192405.8+280352.

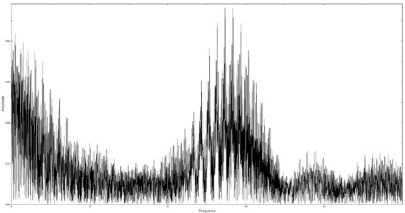

Figure 9. Fourier spectrum of VSX J192405.8+280352.

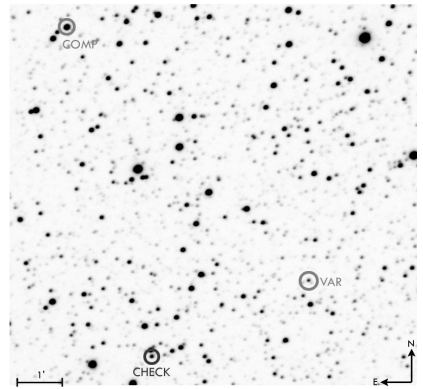

Figure 10. Finding chart of VSX J192252.4+280217.

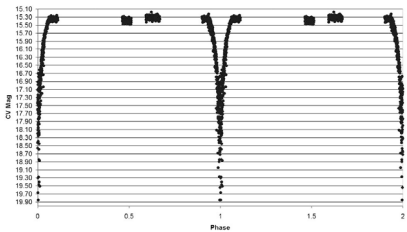

Figure 11a. Phase plot of VSX J192252.4+280217.

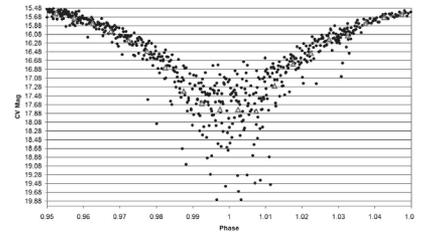

Figure 11b. Phase plot of VSX J192252.4+280217, zoom between phase 0.95 and 1.05. White triangles are binning points.

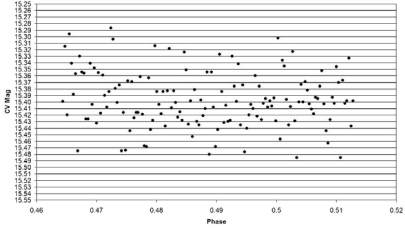

Figure 11c. Phase plot of VSX J192252.4+280217, zoom between phase 0.46 and 0.52. White triangles are binning points. The zoom on Phase between 0.46. 0.52 regards the presence of the secondary minimum, not detected for this variable.

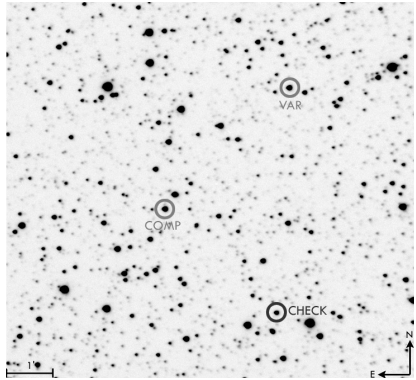

Figure 12. Finding chart of VSX J192251.4+280456.

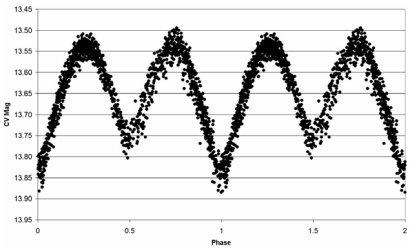

Figure 13. Phase plot of VSX J192251.4+280456.

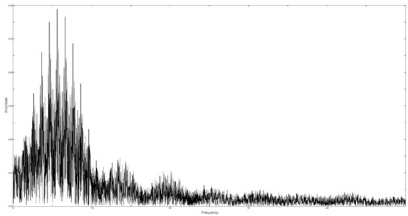

Figure 14. Fourier spectrum of VSX J192251.4+280456.

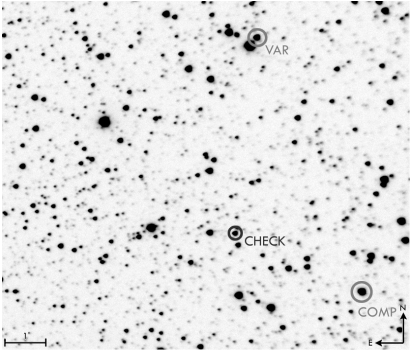

Figure 15. Finding chart of VSX J192226.0+281019.

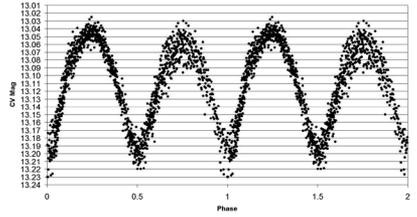

Figure 16. Phase plot of VSX J192226.0+281019.

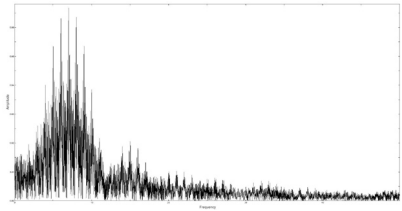

Figure 17. Fourier spectrum of VSX J192226.0+281019.

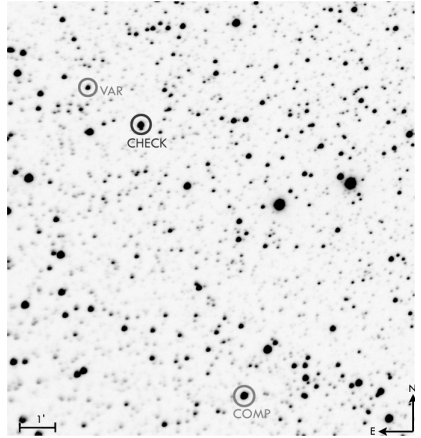

Figure 18. Finding chart of VSX J192524.9+275342.

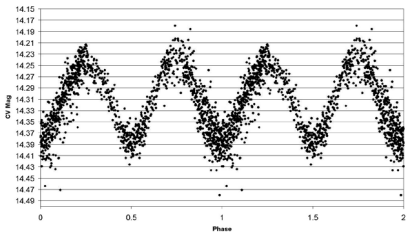

Figure 19. Phase plot of VSX J192524.9+275342.

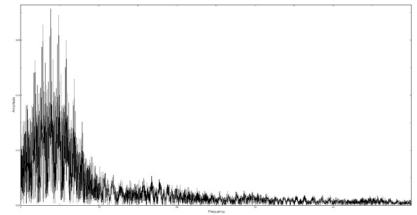

Figure 20. Fourier spectrum of VSX J192524.9+275342.

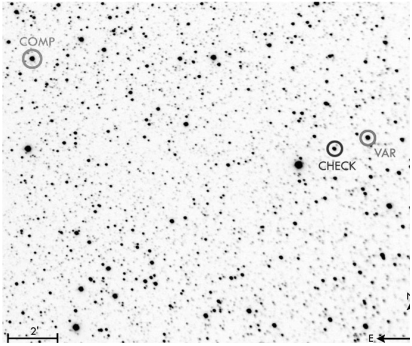

Figure 21. Finding chart of VSX J192220.7+275518.

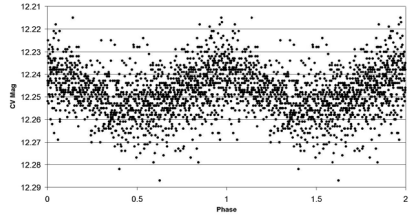

Figure 22. Phase plot of VSX J192220.7+275518.

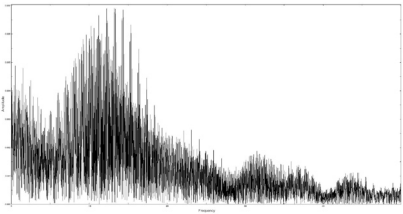

Figure 23. Fourier spectrum of VSX J192220.7+275518.

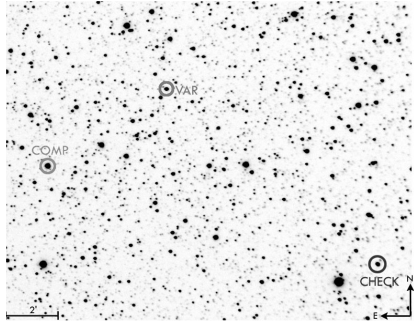

Figure 24. Finding chart of VSX J192304.4+280231.

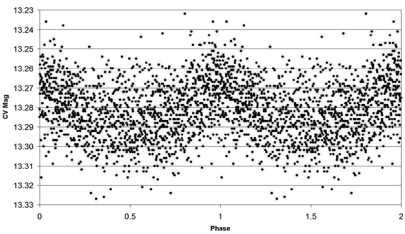

Figure 25. Phase plot of VSX J192304.4+280231.

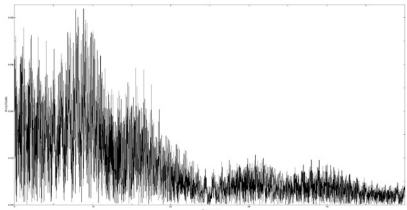

Figure 26. Fourier spectrum of VSX J192304.4+280231.

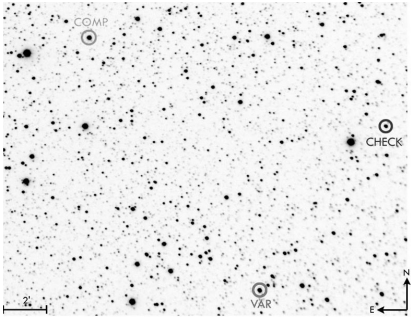

Figure 27. Finder chart of VSX J192255.1+274744.

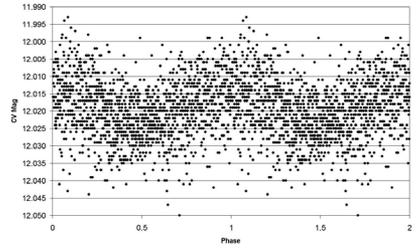

Figure 28. Phase plot of VSX J192255.1+274744.

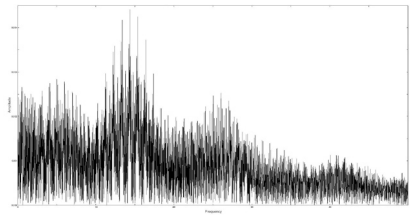

Figure 29. Fourier spectrum of VSX J192255.1+274744.

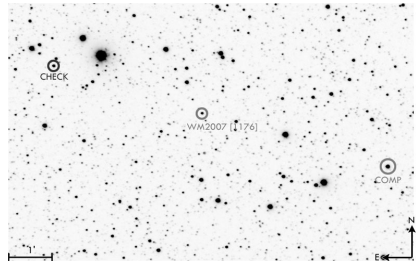

Figure 30. Finding chart of [WM2007] 1176.

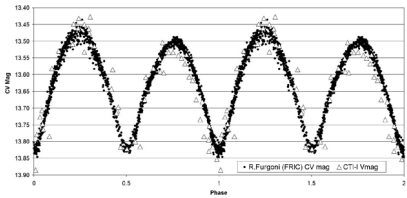

Figure 31. Phase plot of [WM2007] 1176.

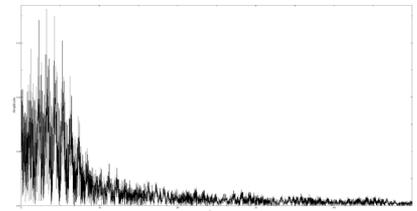

Figure 32. Fourier spectrum of [WM2007] 1176.